\documentclass[11pt]{article} 

\usepackage{type1cm}

\usepackage[latin1]{inputenc}

\usepackage{makeidx}         
\usepackage{graphicx}        

\usepackage{amsmath,amssymb}
\SetSymbolFont{letters}{bold}{OML}{cmm}{b}{it}
\SetSymbolFont{operators}{bold}{OT1}{cmr}{bx}{n}

\usepackage{microtype}
\usepackage{booktabs}

\usepackage{hyperref}

\usepackage{chapterbib}

\usepackage{upgreek}
\usepackage[np]{numprint}
\npdecimalsign{\ensuremath{.}}

\usepackage[thickqspace,cdot,squaren]{SIunits}
\usepackage{desclist}
\usepackage{nicefrac}

\usepackage{eucal}

\usepackage[hang,nooneline]{subfigure}

\usepackage{verbatim}

\usepackage{accents}
\newcommand*{\dt}[1]{%
  \accentset{\mbox{\large .}}{#1}}
\newcommand*{\ddt}[1]{%
  \accentset{\mbox{\large .\hspace{-0.025ex}.}}{#1}}

\usepackage{doi}

\providecommand{\keywords}[1] {
  \small\textit{Keywords:} #1
}

\usepackage{orcidlink}
\usepackage{authblk}

\title{On Common Misconceptions in Classical Vehicle Dynamics}

\author{Massimo Guiggiani\thanks{massimo.guiggiani@unipi.it}\,\,\,\orcidlink{0000-0002-8818-9199}}

\affil{Department of Civil and Industrial Engineering,\\ Universit\`{a} di Pisa, Italy}

\date{}

\begin{document}

\maketitle

\begin{abstract}
Classical vehicle dynamics contains several widely adopted misconceptions that, while intuitively appealing, may lead to inconsistencies when examined under a rigorous mechanical framework. This paper revisits a number of such misconceptions, clarifying their domains of validity and highlighting potential sources of misunderstanding. The aim is not to dismiss established models, but to promote a more precise and scientifically grounded interpretation of key concepts.
\end{abstract}


\section{Introduction}

Classical vehicle dynamics includes several commonly adopted misconceptions that, while appearing reasonable at first glance, may lead to inconsistencies when examined within a rigorous mechanical framework.

These misconceptions often stem from intuitive reasoning, which is a valuable tool in engineering practice. However, difficulties arise when such intuition is extended beyond its domain of validity without sufficient analytical support.

A more rigorous and scientifically grounded perspective is therefore desirable. In this paper, a number of commonly encountered misconceptions are revisited, with the aim of clarifying their limitations and encouraging a more consistent understanding of the underlying mechanics. This discussion is not intended to diminish the value of existing contributions, but rather to complement them by highlighting aspects that may benefit from further clarification.

The topics addressed include:
\begin{itemize}
  \item inertia force and its line of action with respect to the center of mass;
  \item the distinction between center of velocity and center of curvature;
  \item assumptions underlying the single track (bicycle) model;
  \item physical interpretation of lateral acceleration;
  \item interpretation of yaw rate;
  \item interpretation of roll centers and roll axis;
  \item vibration modes, including front/rear mass interpretations and pitch-related aspects.
\end{itemize}

A systematic treatment of vehicle dynamics is presented in \cite{SVD}. A simple, yet rigorous formulation of general dynamics can be found, for example, in \cite{MAMEN}.

\begin{figure}
  \centering
  \includegraphics[width=0.66\textwidth]{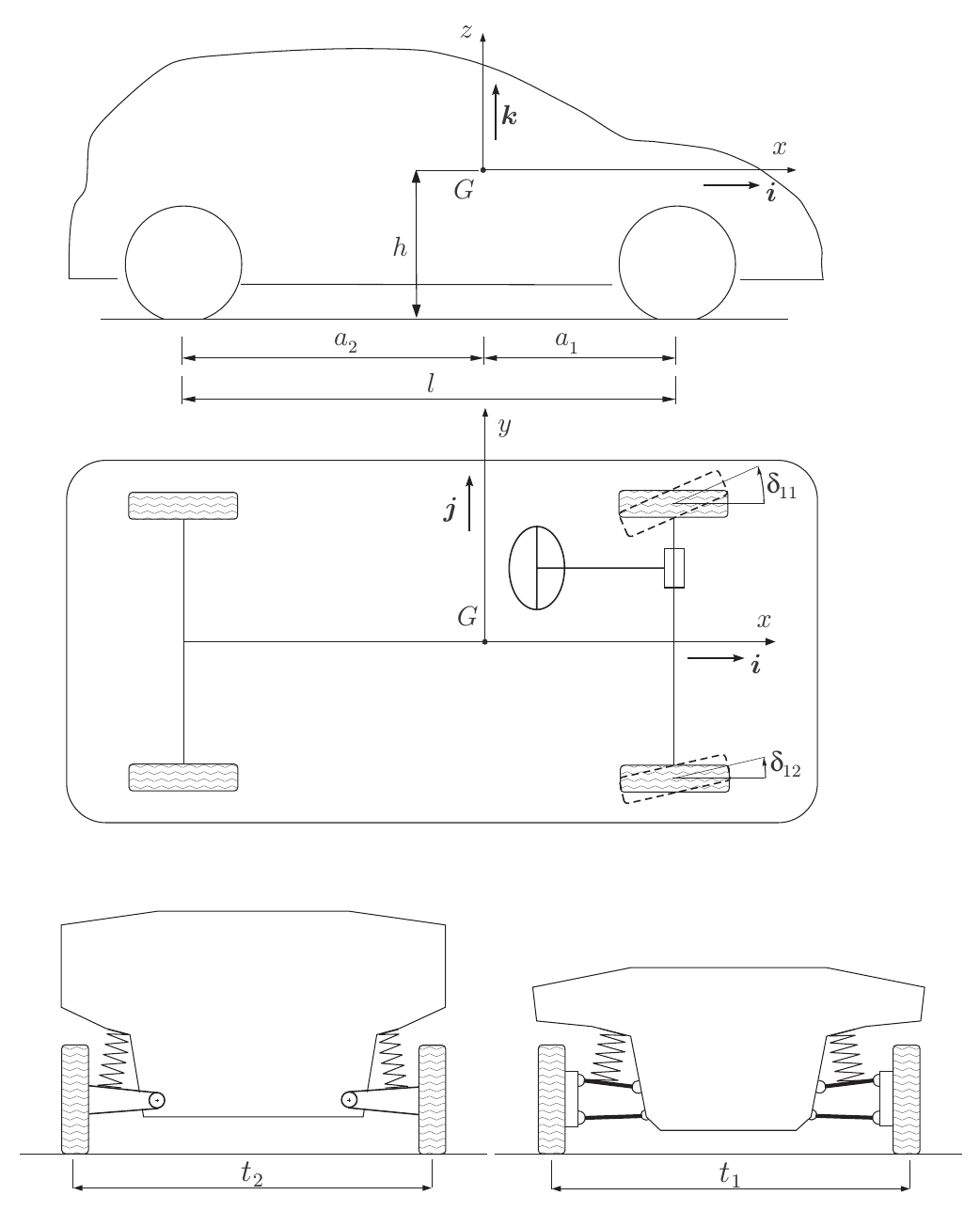}\\
  \caption{Vehicle basic scheme and body-fixed reference system}\label{f:veh_def}
\end{figure}

A vehicle basic scheme and with its body-fixed reference system is shown in Fig.~\ref{f:veh_def}.

\section{Inertia Force not Through $G$}

It is common to find statements such as ``\emph{The centre of mass is the point where all of the mass of the car can be considered to be concentrated}" \cite[p.~{4}]{SEWARD} or ``\emph{the centrifugal force ALWAYS acts at the centre of gravity}" \cite[pp.~{39, 41, 132}]{PUTZ}. Similar statements can be found in
\cite[p.~{223}]{SCHRAMM} and
\cite[p.~{149}]{WILLIAMS}.

These statements are correct only within a limited context. It is indeed true that the \emph{resultant inertia force} $\mathbf{R}$ is equal to the mass $m$ multiplied by the acceleration $\mathbf{a}_G$ of the center of mass $G$:
\begin{equation}\label{e:1}
  m\,\mathbf{a}_G = \mathbf{R}
\end{equation}
However, in general, inertia forces also produce a \emph{resultant moment} $\mathbf{M}_G$ with respect to $G$.
Therefore, if one seeks an equivalent representation consisting of a single force with zero resultant moment, the corresponding line of action of $\mathbf{R}$ does not, in general, pass through $G$.

Only under steady-state conditions does the line of action of $\mathbf{R}$ coincide with the center of mass \cite[p.~{9}]{PACEJKA02}, \cite[p.~409]{SVD}.

The interpretation of the vehicle as a point mass located at $G$, while useful in specific contexts, is therefore an oversimplification that may lead to inaccuracies when applied beyond its domain of validity \cite[p.~ 111]{WILLIAMS}. Moreover, reliance on equation \eqref{e:1} alone may obscure the role of the associated moment $\mathbf{M}_G$.


\begin{figure}[h]
\centering
  \includegraphics[width=0.617\textwidth]{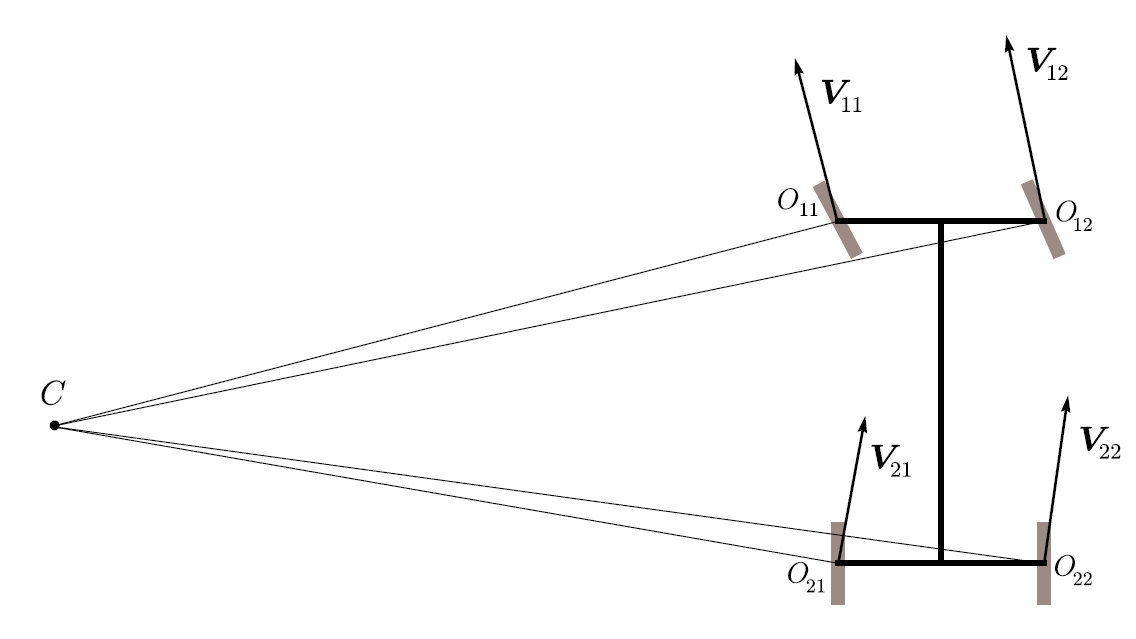}
  \caption{Center of velocity}\label{f:cv}
\end{figure}

\section{Center of Velocity not Being the Center of Curvature}

A vehicle in planar motion, like any rigid body, has an \emph{instantaneous center of velocity} $C$, as shown in Fig.~\ref{f:cv}, defined as the point with zero velocity. Accordingly, the \emph{velocity field} of the rigid body can be interpreted as a pure instantaneous rotation about $C$.

\begin{figure}[h]
\centering
  \includegraphics[width=0.617\textwidth]{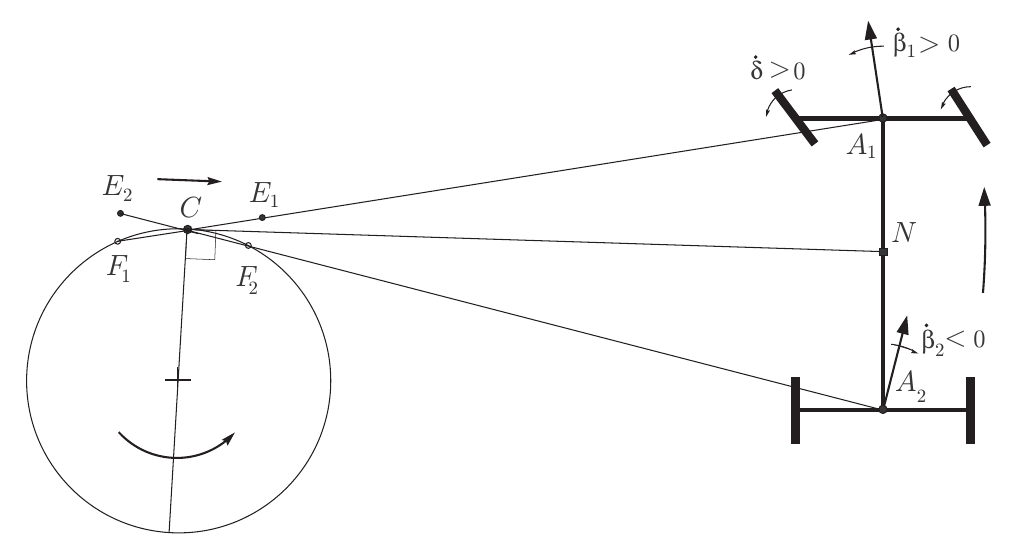}
  \caption{Radii of curvature $E_1A_1$ and $E_2A_2$ of the trajectories of points $A_1$ and $A_2$, respectively, of a vehicle entering a left turn}\label{f:goodentering}
\end{figure}

\begin{figure}
\centering
  \includegraphics[width=0.617\textwidth]{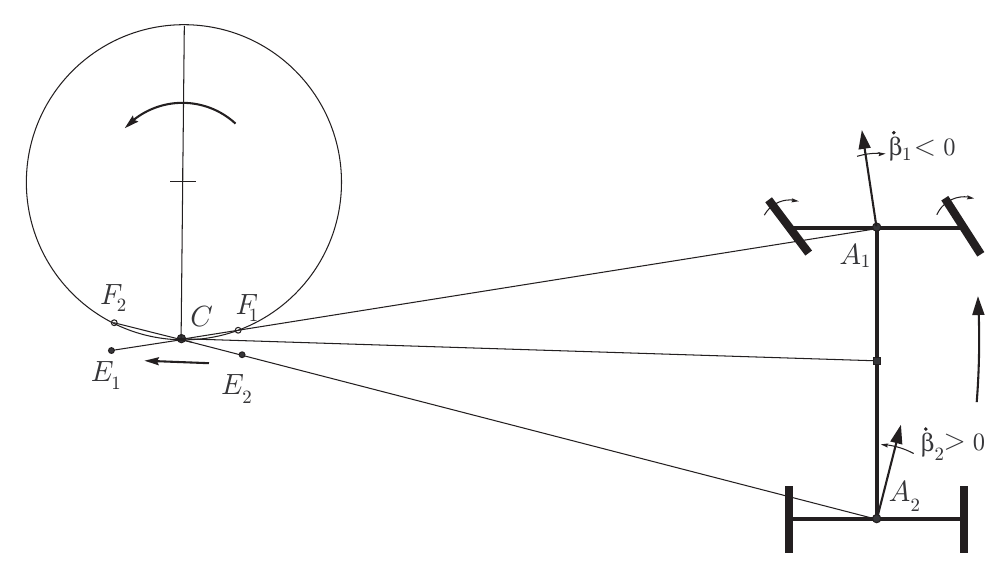}
  \caption{Radii of curvature of a vehicle exiting a left turn}\label{f:goodexiting}
\end{figure}

However, this property does not extend to the \emph{acceleration field}. In general, the acceleration $\mathbf{a}_C$ of point $C$ is not zero \cite[p.~{72}]{SVD}. For this reason, $C$ is more precisely referred to as the instantaneous center of zero velocity.

Alternative names for $C$, such as ``turning point" \cite[p.~{127}]{MILLIKEN}, ``instantaneous center of motion" \cite[p.~{317}]{KIENCKE}, or ``center of rotation" \cite{JAZAR4}, may be ambiguous if interpreted beyond their precise kinematic meaning.

The identification of the center of velocity $C$ with a center of curvature (\cite[p.~{92}]{FROMMIG}, \cite[p.~{172}]{PAUWELUSSEN}) is only valid under steady-state conditions. In general transient motion, at a given instant, there exists a unique center of velocity, whereas each point of the vehicle follows its own trajectory, characterized by its own center of curvature. Therefore, infinitely many centers of curvature coexist.

Statements such as ``The center of curvature of the road is supposed to be the turning center of the car at the instant of consideration" \cite[p.~{339}]{JAZARAVD} should therefore be interpreted with caution.

In Fig.~\ref{f:goodentering} from \cite{SVD}, points $E_1$ and $E_2$ represent the centers of curvature of the trajectories of points $A_1$ and $A_2$, respectively. These points do not, in general, coincide with the center of velocity $C$. This situation is typical when entering a curve, i.e., when the steering angle is increasing. In this case, the front axle midpoint has a shorter radius of curvature than the rear axle midpoint, i.e.,
\begin{equation}
|E_1A_1| < |E_2A_2|
\end{equation}

Conversely, when exiting a curve, the situation is reversed, as shown in Fig.~\ref{f:goodexiting}. In both figures it is also shown the \emph{inflection circle} \cite[Sect.~5.4.1]{SVD}. We recall the noteworthy relationship
\begin{equation}
  |A_i C|^2 = |A_i E_i||A_i F_i|
\end{equation}


\section{Assumptions of the Single Track (Bicycle) Model}

The term ``bicycle model'' is widely used in the literature, although it may be somewhat misleading, as it does not correspond to a physical model of a bicycle. Similarly, the expression ``two-wheel model'' may also lead to ambiguity \cite[p.~{17}]{PACEJKA02}.

\begin{figure}[h]
\centering
  \includegraphics[width=0.75\textwidth]{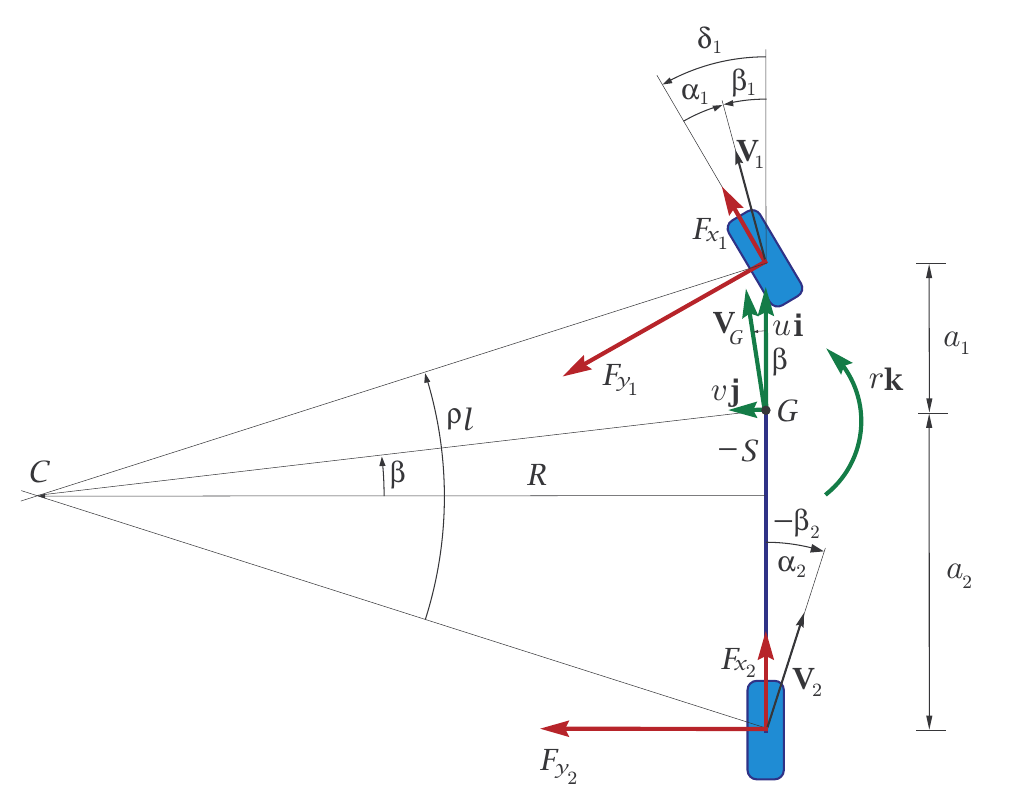}
  \caption{Single track model}\label{f:fmonotr}
\end{figure}

As discussed in \cite[Chap.~{7}]{SVD}, the model represents a vehicle with four wheels, despite its simplified graphical representation as a single track, such as that shown in Fig.~\ref{f:fmonotr}.

As correctly noted by Pacejka \cite[p.~{17}]{PACEJKA02}, the single track model is capable of accounting for effects such as lateral load transfer and body roll. A detailed discussion of its features can be found in \cite[Sect.~{7.5}]{SVD}, including extensions of the axle characteristics that incorporate also  toe in/out, camber, and roll steer.

Moreover, the single track model is often introduced in a highly simplified manner, without explicitly stating the underlying assumptions \cite[p.~{126}]{MILLIKEN}, \cite[p.~{54}]{ABE}, \cite[p.~{84}]{BARBIERI}, \cite[p.~{140}]{BLUNDELL}, \cite[p.~{93}]{FROMMIG}, \cite[p.~{583}]{JAZAR3}, \cite[p.~{115}]{JAZARAVD}, \cite[p.~{170}]{MEYWERK}, \cite[p.~{224}]{SCHRAMM}, leading often to incorrect statements.

For instance, assumptions such as placing the center of mass $G$ at road level \cite[p.~{170}]{MEYWERK}, or considering equal lateral forces on the left and right tires \cite[p.~{53}]{ABE}, are not required by the model and do not reflect typical vehicle behavior.

Similarly, treating the vehicle as a point mass concentrated at $G$ \cite[p.~{223}]{SCHRAMM} is an additional simplification that is not inherent to the formulation and may limit its applicability.

The graphical representation of the single track model, as in Fig.~\ref{f:fmonotr}, should therefore be interpreted as a convenient abstraction for deriving the governing equations of a four-wheel vehicle, such as that shown in Fig.~\ref{f:veic}. In this sense, the model does not imply a reduction of the physical system to a single track, but rather a simplified representation of its dynamics \cite[p.~{272}]{SVD}. The single track model is not really single track!


\section{Lateral Acceleration}

\begin{figure}[h]
\centering
  \includegraphics[width=0.4\textwidth]{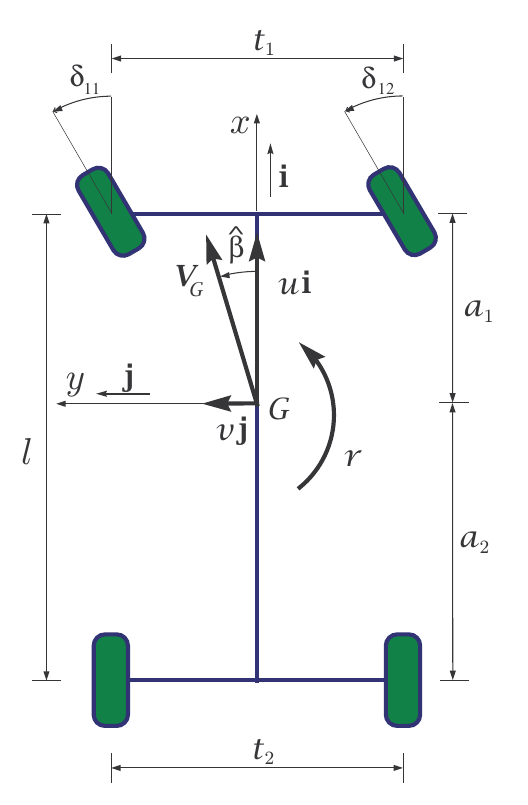}
  \caption{Vehicle axis system and global kinematics of a vehicle in planar motion}\label{f:veic}
\end{figure}

Let us consider again the vehicle as a rigid body in planar motion (Fig.~\ref{f:veic}).

Let $\mathbf{V\!\!\,}_G = u \,\mathbf{i} + v \,\mathbf{j}$ be the velocity of $G$. The yaw rate is $\dt{\psi} \,\mathbf{k} = r \,\mathbf{k}$. The acceleration of $G$ is \cite[p.~{75}]{SVD}
\begin{equation}\label{e:acc}
\begin{split}
  a_x &= \dt{u} - v r \\
  a_y &= u r + \dt{v}
\end{split}
\end{equation}

Although \eqref{e:acc} is elementary, its interpretation is sometimes a source of misunderstanding.

A first issue concerns the notation. In some references, the lateral acceleration $a_y$ is denoted as $\ddt{y}$ \cite[p.~{77}]{ABE}, \cite[pp.~{89, 95}]{FROMMIG}, \cite[p.~{28}]{RAJAMANI}. This notation may be misleading, since $a_y$ is not, in general, the second time derivative of a single scalar coordinate, but rather the projection of the acceleration vector onto the body-fixed $y$-axis.

A second point concerns the physical interpretation of the terms in $a_y$. Expressions such as ``the first term $u r$ represents a centrifugal contribution, while the second represents a direct lateral acceleration'' \cite[p.~{146}]{MILLIKEN} (see also \cite[p.~{130}]{BALKWILL}) may erroneously suggest that these terms correspond to distinct physical effects.

Similarly, statements such as ``Lateral forces on the tires from the road at the front and the rear of the vehicle are opposed to the centrifugal force and the difference accelerates the car laterally'' \cite[p.~{170}]{WILLIAMS} rely on a comparable interpretation.

\begin{figure}[h]
   \centering
    \includegraphics[width=0.9\textwidth]{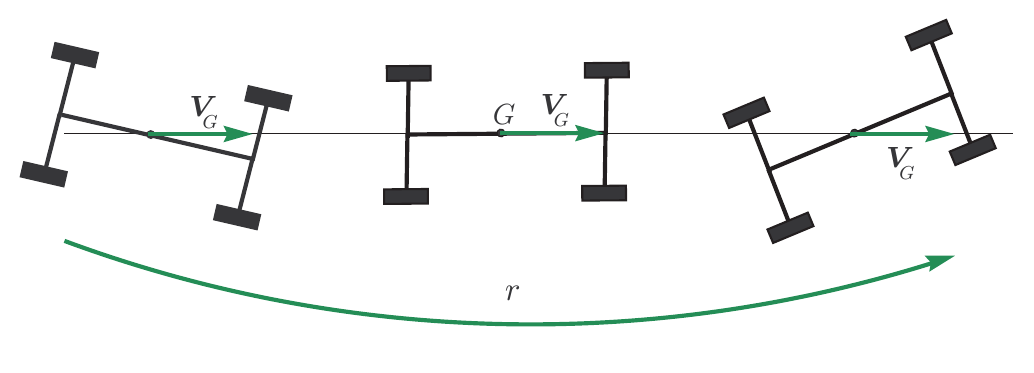}
 \caption{Motion with zero acceleration $\mathbf{a}_G$, although $u$, $v$, $r$, $\protect \dt{u}$, $\protect \dt{v}$ and $\protect \dt{r}$ are all non-zero}
 \label{f:retta}
\end{figure}

As a matter of fact, Eq.~\eqref{e:acc} is a kinematic identity, and the terms $u r$ and $\dt{v}$ arise from the time derivative of the velocity components in a rotating reference frame. As such, they do not represent independent physical contributions, but components of the same acceleration vector expressed in body-fixed coordinates.
This point can be clarified by considering the example shown in Fig.~\ref{f:retta}.

In this case, the center of mass $G$ moves with constant velocity $\mathbf{V\!\!\,}_G$ along a straight line, while the vehicle has a non-zero yaw rate $r$. Consequently, $a_y = 0$, despite the fact that both $u r \neq 0$ and $\dt{v} \neq 0$.

This example shows that the individual terms in Eq.\,\eqref{e:acc} cannot be interpreted independently as distinct physical effects. Only their sum has a direct physical meaning, namely the lateral component of the acceleration of $G$.

Therefore, decompositions of $a_y$ into ``centrifugal'' and ``lateral'' contributions should be interpreted with caution, as they do not correspond to a general physical separation of effects.


\section{One Yaw Rate, not Two}

Let us consider the vehicle as a rigid body in planar motion (Fig.~\ref{f:veic}). The vehicle has a yaw rate $\dt{\psi} = r$, which is a property of the rigid body as a whole.

It is important to emphasize that the yaw rate represents the angular velocity of the body, and not a rotation about a particular point. In planar rigid-body kinematics, the angular velocity is uniquely defined and does not depend on the choice of reference point.

In the literature, however, different interpretations of yaw rate are sometimes introduced. The following examples illustrate some formulations that may lead to ambiguity:
\begin{itemize}
  \item ``The yaw rate consists of two contributions: the first is the rotation of the vehicle center of gravity around the center of curvature, the second is the rotation of the vehicle around the center of gravity (rotation on itself)" \cite[p.~{208}]{BARBIERI}.
  \item ``The second approach is to obtain the vehicle movement as an orbit around the instantaneous center with angular velocity $\dt{\psi}$ and $\dt{\beta}$ and then calculate the radius to each individual wheel"  \cite[p.~{305}]{KIENCKE}.
  \item ``Yaw rate, $r$, is the angular velocity of the vehicle around a vertical axis passing through the CG."  \cite[p.~{144}]{MILLIKEN}.
  \item ``Yaw rate about the turning center" \cite[p.~{221}]{MILLIKEN}.
\end{itemize}

These formulations often arise from attempts to interpret the motion in terms of trajectories or reference points (e.g., center of curvature or instantaneous center). While such constructions may be useful for geometric visualization, they should not be interpreted as defining distinct yaw rates.

In rigid-body planar motion, there is a single angular velocity vector associated with the body. This angular velocity is independent of the reference point and fully characterizes the rotational motion.

The yaw rate is thus a unique quantity, and interpretations involving multiple yaw rates should be understood as alternative descriptions of the same kinematic variable, rather than independent physical entities.


\section{Roll Centers and Roll Axis}

As discussed by Dixon in \cite{DIXONROLL}:
``Perhaps even more importantly, the very concept of the body rolling about its roll-axis is a rather dubious one, and without the constant-track approximation it becomes especially difficult to give a sound physical interpretation to the \emph{kinematic roll-centre}.''

This observation highlights the need for a careful interpretation of roll-related concepts. In particular, the notion of a kinematic roll axis should be treated with caution. As discussed in \cite[Ch.~{9}]{SVD}, the vehicle is a deformable system supported by suspensions and tires, and its motion cannot, in general, be reduced to a pure rotation about a fixed axis.

Consequently, interpreting the roll axis as a kinematic entity analogous to a revolute joint \cite[p.~{333}]{BLUNDELL} may lead to conceptual inconsistencies. Similarly, statements implying that the sprung mass rolls about a well-defined roll center should be understood as approximations valid only under specific assumptions.

Still following Dixon \cite{DIXONROLL}: ``However, as innumerable examples in the literature show, the usual application of the roll-centre is in evaluating the lateral load transfer during cornering.''
In this context, the concept of \emph{force roll center} becomes more appropriate.

The \emph{force roll center} is addressed in \cite[Ch.~{3}]{SVD} where it is called \emph{no-roll center} to highlight its main feature: a lateral force applied at it does not provoke roll of the vehicle body due to the suspensions (but only roll due to the tires).

\begin{figure}[t]
\centering
  \includegraphics[width=0.7\textwidth]{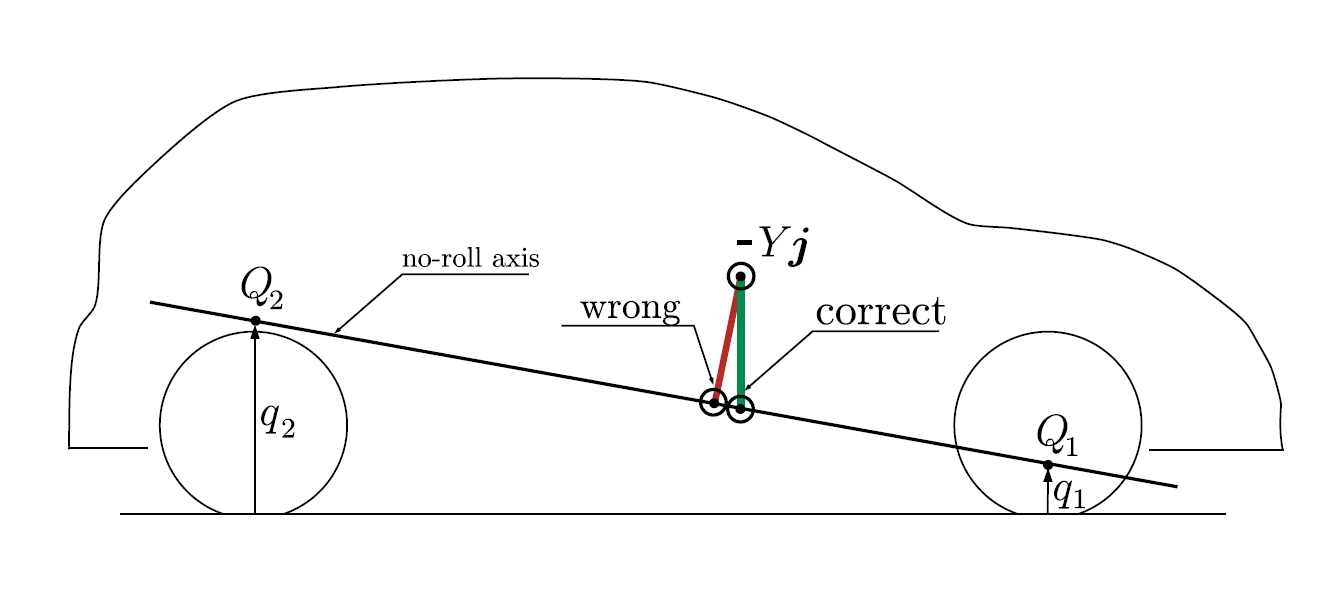}
  \caption{Evaluation of the roll moment based on force distribution}\label{f:wrong}
\end{figure}

The straight line connecting the front and rear no-roll centers is commonly referred to as the roll axis. However, it should be emphasized that this line does not represent a physical axis of rotation. A better name would be \emph{no-roll axis}.

Difficulties may arise when roll moments are evaluated with respect to this axis. For example, statements such as ``The centrifugal force acting at the center of gravity produces a rolling moment around the roll axis resulting in a constant roll angle'' \cite[p.~{169}]{ABE} rely on an interpretation that may not be generally valid. As illustrated in Fig.~\ref{f:wrong} from \cite[p.~{122}]{SVD}, the relevant moment is not associated with a rotation about the roll axis, but should instead be evaluated directly from the distribution of forces and their lines of action. Similar considerations apply to formulations that employ distances measured from the roll axis to compute roll moments \cite[p.~{8}]{PACEJKA02}, \cite[p.~{681}]{MILLIKEN}; see also \cite[p.~{67, 133}]{SEWARD}, \cite[p.~{61}]{BARBIERI}.

A comprehensive analysis of roll-related phenomena, including the role of suspension geometry, tire compliance, and jacking forces can be found in \cite[Sect.~{3.10}]{SVD}.


\section{Vibration Modes (Ride)}

A typical car has four wheels, arranged in two axles (front and rear), often equipped with independent suspensions.

\begin{figure}[h]
\centering
  \includegraphics[width=0.3\textwidth]{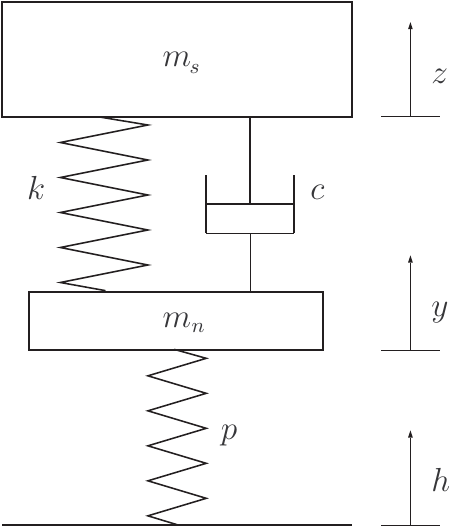}
  \caption{Quarter-car model}\label{f:mono}
\end{figure}

\begin{figure}[h]
\centering
  \includegraphics[width=0.6\textwidth]{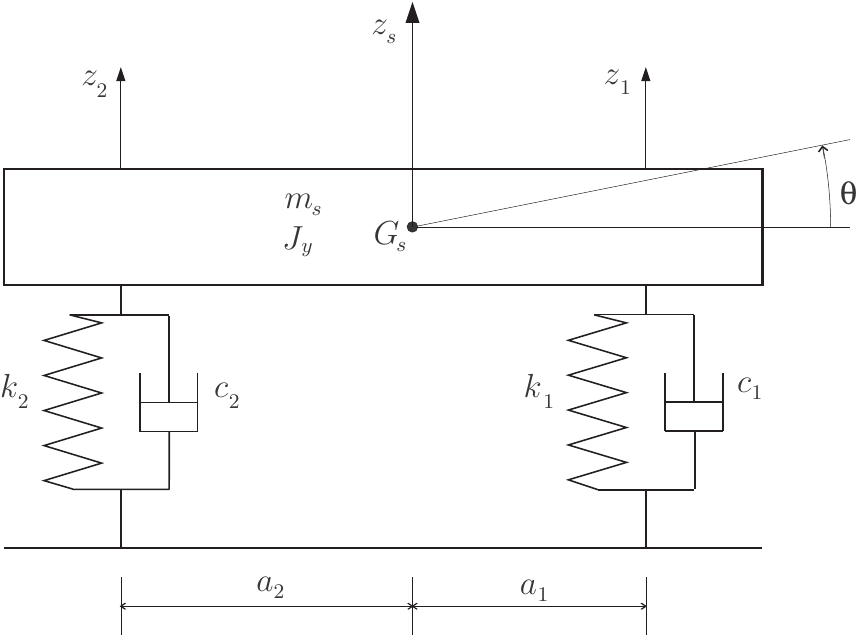}
  \caption{Bounce--pitch model}\label{f:dof2}
\end{figure}

Two simple models are usually introduced to investigate ride behavior:
\begin{itemize}
  \item the quarter-car model, shown in Fig.~\ref{f:mono}, with a translating sprung mass $m_s$ and a translating unsprung mass $m_n$;
  \item the bounce--pitch model, shown in Fig.~\ref{f:dof2}, with a rigid body of mass $m_s$ and moment of inertia $J_y$ with respect to $G_s$;
\end{itemize}
and this is fully appropriate. Both models have two degrees of freedom and include linear springs and dampers.

A linear system with two degrees of freedom has two natural modes, hence two natural frequencies and two damping ratios. Each natural mode involves both degrees of freedom. This is a classical result.

Regarding the quarter-car model (Fig.~\ref{f:mono}), statements such as ``Because tyres act as springs, the unsprung mass also has a natural frequency, bouncing between the tyre spring and the suspension spring" \cite[p.~{14}]{EDGAR} may suggest that each mass has its own independent natural frequency. Strictly speaking, this interpretation is not correct.

In a road car, one typically has $m_s \gg m_n$ and $k \ll p$. As a consequence, in the first mode the motion of $m_s$ is more pronounced than that of $m_n$, whereas in the second mode the opposite occurs. Nevertheless, both degrees of freedom participate in both modes. Similar remarks apply to statements such as those in \cite[p.~{95}]{SEWARD}.

Moreover, in race cars $m_s > m_n$ and $k < p$, but usually to a lesser extent than in road cars, leading to a stronger interaction between the two masses.

A similar issue may arise for the bounce--pitch model (Fig.~\ref{f:dof2}). For instance, the sentence ``Most cars have different front and rear natural frequencies" \cite[p.~{11}]{EDGAR} may suggest that the system shown in Fig.~\ref{f:dof2} behaves as two independent one-degree-of-freedom systems. Again, this interpretation is not generally correct.

Following a similar line of reasoning, some authors introduce a front mass and a rear mass \cite[p.~{85}]{BARBIERI}. While such a description may be heuristically appealing in some contexts, it is not required by the model and may obscure its coupled nature.

The bounce--pitch model is also often presented without dampers \cite[p.~{172}]{GILLESPIE}, as if their role were negligible. In an actual vehicle, however, dampers are clearly not negligible. As discussed in \cite[Sect.~{10.6.2}]{SVD}, the undamped vibrational behavior may still provide useful information for the damped system, provided that in the car \emph{proportional damping} is employed, i.e., for the system in Fig.~\ref{f:dof2}
\begin{equation}
  \frac{c_1}{k_1} = \frac{c_2}{k_2}
\end{equation}
This condition, however, is seldom stated explicitly. The problem is that without proportional damping we do not have a bounce mode and a pitch mode, and the ride behavior of the car becomes uncomfortable.

More serious difficulties arise when the idea of decoupling the front and rear axles, together with the introduction of front and rear masses, is extended to the evaluation of load transfers \cite[p.~{389}]{KIENCKE}. With the notation of Figg.~\ref{f:veh_def} and \ref{f:veic}, Equation (9.52) in \cite{KIENCKE} for the vertical load $Z_{12}$ on the right front wheel reads
\begin{equation}
  \begin{split}
    Z_{12} &= m \left( \frac{a_2 g}{l}-\frac{h a_x}{l}\right)
    \left[\frac{1}{2} + \frac{h a_y}{t_1 g} \right]
    \\
      &=
      m g\frac{a_2}{2l}- m a_x\frac{h}{2l}+m a_y \frac{a_2 h}{l t_1} - m a_x a_y\frac{h^2}{l t_1 g}
  \end{split}
\end{equation}

The appearance of a term proportional to the product $a_x a_y$ is a clear indication that the derivation is based on assumptions that are not mechanically consistent.



\section{Conclusions}

This paper has revisited several commonly used interpretations in classical vehicle dynamics, emphasizing their domains of validity and highlighting potential sources of misunderstanding.

The analysis shows that many of these issues arise from extending simplified or steady-state interpretations beyond their applicable range, or from attributing physical meaning to quantities that originate from kinematic relations or modeling abstractions.

A consistent treatment of vehicle dynamics requires careful distinction between kinematics and dynamics, as well as a clear identification of the assumptions underlying each model. When these aspects are properly accounted for, many apparent inconsistencies can be resolved.

The aim of this work is not to question the usefulness of simplified models, but to promote a more rigorous interpretation of their results. It is hoped that this perspective may contribute to a clearer and more consistent understanding of fundamental concepts in vehicle dynamics.


\end{document}